# Using Identity-Based Cryptography in Mobile Applications

V. Mora-Afonso, P. Caballero-Gil

Department of Statistics, O.R. and Computing. University of La Laguna. Spain.
alu3966@etsii.ull.es, pcaballe@ull.es

**Abstract.** This work includes a review of two cases study of mobile applications that use Identity-Based Cryptography (IBC) to protect communications. It also describes a proposal of a new mobile application that combines the use of IBC for Wi-Fi or Bluetooth communication between smartphones, with the promising Near Field Communication (NFC) technology for secure authentication. The proposed scheme involves NFC pairing to establish as public key a piece of information linked to the device, such as the phone number, so that this information is then used in an IBC scheme for peer-to-peer communication. This is a work in progress, so the implementation of a prototype based on smartphones is still being improved.

**Keywords:** Identity-Based Cryptography, Mobile Application, Near Field Communication

## 1 Introduction

The use of smartphones has been increasing rapidly worldwide for the last years so that they have overtaken computers. Moreover, the figures say that this tendency is to be accelerated in the next future [1]. Smartphones are nowadays used for everything: checking email, social networking, paying tickets, communication between peers, data sharing, etc. However, in general they have lower computing capabilities than a standard computer, as well as power and battery limitations. Thus, every operation or computation in a smartphone must be implemented taking into account these constraints, maximizing efficiency and memory usage in order to avoid possible problems. In spite of these difficulties, still secure communication mechanisms and protocols need to be provided in different layers and applications in order to offer security to end users in a transparent way so that they can find functional mobile applications both user-friendly and robust.

The so-called Identity-Based Cryptography (IBC) may be seen as an evolution of Public-Key Cryptography (PKC) because IBC gets rid of most of the problems of traditional Public-Key Infrastructures (PKIs), as they are usually related to certificates. Furthermore, it provides the same security level, but using shorter encryption keys and more efficient algorithms. In the past decade, IBC has been subject of intensive study, and many different encryption, signature, key agreement and signcryption schemes have been proposed [2, 3, 4].

Near Field Communication (NFC) is a short-range high frequency wireless communication technology [5] that enables simple and safe interactions between electronic devices at a few centimeters, allowing consumers to perform contactless transactions, access digital content, and connect electronic devices with a single touch.

This paper aims to review some mobile applications based on mechanisms and protocols that use IBC to provide security, and to propose a new NFC-based mobile application for Bluetooth/Wi-Fi pairing for peer-to-peer communication whose security is based on IBC. The characteristics of these schemes perfectly fit with the demand and requirements of not only smartphones but also backend applications and servers in terms of simplicity, security, efficiency and scalability. In particular, two known proposals are here analyzed and discussed. The first one aims to protect information sharing among mobile devices in dynamic groups whereas the second one proposes a secure protocol for communications in social networks, both using Identity Based Cryptography as the main security mechanism.

The remainder of this paper is organized as follows. In Section 2 a brief introduction to the cryptographic primitives used throughout the document is included. Then, Section 3 reviews the two mobile applications chosen for being studied. Section 4 presents the proposed scheme and discusses its design, security and implementation. Finally, Section 5 concludes the paper, including some future work.

## 2    Background

Identity-Based Cryptography is a type of public-key cryptography where a public piece of information linked to a node is used as its public key. Such information may be an e-mail address, domain name, physical IP address, phone number, or a combination of any of them. Shamir described in [6] the first implementation of IBC. In particular, such a proposal is an identity-based signature that allows verifying digital signatures by using only public information such as the user's identifier. Shamir also proposed identity-based encryption, which appeared particularly attractive since there was no need to acquire an identity's public key prior to encryption. However, he was unable to come up with a concrete solution, and identity-based encryption remained an open problem for many years.

### 3.1 Identity-Based Encryption

Identity-Based Encryption (IBE) is a public cryptographic scheme where any piece of information linked to an identity can act as a valid public key. A Trusted Third Party (TTP) server, the so-called Private Key Generator (PKG), first stores a secret master key used to generate a set of public parameters and the corresponding users' private keys. After a user's registration, it receives the public parameters and its private key. Then, a secure communication channel can be established without involving the PKG.

Boneh and Franklin proposed in [2] the first provable secure IBE scheme. Its security is based on the hardness of the Bilinear Diffie-Hellman Problem [7].

IBE schemes are usually divided into four main steps:

1. Setup: This phase is executed by the PKG just once in order to create the whole IBE environment. The master key is kept secret and used to obtain users' private keys, while the remaining system parameters are made public.
2. Extract: The PKG performs this phase when a user requests a private key. The verification of the authenticity of the requestor and the secure transport of the private key are problems with which IBE protocols do not deal.
3. Encrypt: This step is run by any user who wants to encrypt a message M in order to send the encrypted message C to a user whose public identity is ID.
4. Decrypt: Any user that receives an encrypted message C runs this phase to decrypt it using its private key d and recovering the original message M.

### 3.2 Hierarchical Identity-Based Encryption

Hierarchical Identity-Based Encryption (HIBE) may be seen as a generalization of IBE that reflects an organizational hierarchy. Thus, it is an organizational-hierarchy oriented generalization of IBE [8][9]. It lets a root PKG distribute the workload by delegating private key generation and identity authentication to lower-level PKGs, improving scalability. In this way, a possible disclosure of a domain PKG's secret does not compromise the secrets of higher-level PKGs.

### 3.3 Attribute-Based Encryption

Attribute-Based Encryption (ABE) is a type of public-key encryption in which a user's public key or a ciphertext is associated with a set of attributes (e.g. the country the user lives in, the kind of subscription the user has, etc.). This scheme provides complete access control on encrypted data by setting up policies and attributes on ciphertexts or keys. Two different types of ABE can be distinguished: Ciphertext Policies ABE (CP-ABE) and Key Policies ABE (KP-ABE). In CP-ABE the decryption of a ciphertext is only possible if the set of attributes of the user key matches the attributes of the ciphertext [10]. On the other hand, in KP-ABE, ciphertexts are labeled with sets of attributes, and private keys are associated with access structures that control which ciphertexts a user is able to decrypt [11].

## 3    Two Cases Study

Through the following description of two-cases study, it is demonstrated that the implementation of Identity-Based Cryptosystems, which have many advantages over traditional PKIs, is possible, practical and efficient not only in smartphones but in everyday applications such as social networks, securing users communications transparently.

### 3.1 Secure Information Sharing in Mobile Groups

The work [12] proposes an application for mobile devices that can be used to share information in groups thanks to a combination of HIBE and ABE, and describes its implementation for the Android operating system. In particular, it describes a protocol called Trusted Group-based Information Sharing (TGIS) to establish group-based trust relationships in order to share information within a group or among groups. TGIS defines a cryptographic access control scheme to provide secure communication among mobile devices of collaborators who belong to the same or different organizations, and/or with different ranks or privileges. It assumes that each device belongs to an organization that has implemented some hierarchical system, what allows deploying credentials that can be used as public keys to distribute group keys. On the other hand, for controlling information shared within a group, secure access control is based on attributes. With this application, groups can be created dynamically and by any user, who acts as group leader in charge of generating the corresponding private key for each group member. Thus, since these members can be from different organizations, the group leader in fact acts as group PKG.

An interesting use case of this application is in emergency situations such as an earthquake or a volcanic eruption because in this kind of environments many groups from very different types need to communicate and collaborate in order to reach a common goal. Police officers, emergency services and firefighters may have to share information, data, locations, etc. without making this knowledge publicly available.

Going into detail, TGIS protocol is divided into six phases.

1. The first stage, named Domain Setup, is when each user registers its device with an identity in its organization's hierarchical domain, in order to receive a HIBE private key from a PKG. The PKG makes public a set of parameters used to generate HIBE public keys from user identities, and keeps secret the master private key used to generate each user's HIBE private key.
2. The second stage, named Group Setup, starts when a group has to be created because peers need to share information. Then, one of these users becomes the group leader in order to generate a public/private key pair and a set of group attributes for the group and signs with its HIBE private key the generated public key and the group attributes so that other users can verify this message using the group leader's HIBE identity.
3. Once a group has been created, the third phase, named User Enrollment, begins. Whenever a user wants to join a group, the leader decides what privileges it will have, assigns privileges to the group set of attributes, creates the group private key for the new user and sends it securely using the new user's HIBE identity.
4. When a group has been created, information can be shared among members, so the fourth phase, called Intra-Group Communication, can be executed. Users can share data and decrypt them complying with the group access policies defined in the group setup phase, so that only group members with the proper permissions can access the shared information. When a user encrypts a message, it defines which attributes can decrypt it using the group public key and an access tree describing these policies.

5. Communication among different groups can be achieved through the fifth phase called Inter-Group Collaboration, where every group leader makes available the group public key and the set of attributes.
6. Finally, in order to achieve message authentication in the sixth defined phase, every message exchanged within a group or among groups must be signed using an identity-based signature scheme.

In order to generate trust in dynamic collaborative groups for exchanging information between mobile devices, the described work proposed a distributed security protocol based on HIBE to provide flexible and secure access control. Such a proposal was implemented and evaluated in Android phones, what proves that this type of cryptography may be used in different applications.

### 3.2 Secure Smartphone-Based Social Networking

Given the use of social networks nowadays and the lack of secure mechanisms to allow communication among users of different social providers, if a user's account gets compromised, all their contacts get indirectly compromised and could become victims of different attacks such as phishing. There are a few applications [13, 14, 15] that prevent users from being scammed, provide access control, etc., but usually current social networks security schemes assume third parties are honest-but-curious and so they are not as scalable as current social networks demand. In [16] a security-transparent smartphone-based solution to this problem is presented. This approach uses IBC as mechanism to protect data, which can be applied in any third-party social network provider. An Android implementation of such a scheme is presented as a mobile application using WeChat [17].

In every social network scheme usually many different parts are involved. They are: a social network provider, a PKG to which users connect in order to get their private keys based on their public ID (mostly phone number) and users, who send and/or receive data from other users through their smartphone (in which they can perform reasonable computing operations efficiently). In order to gain access control, before sending data, the user must encrypt them using the receiver's identity. On the other side of the channel, when the user receives data, they must get their private key from the PKG in order to be able to decrypt data.

Given that an adversary can compromise a server or a PKG, in the security model of this scheme, certain assumptions are made. A compromised social network provider might tamper the communication, inserting, deleting or impersonating any user. However, a compromised PKG is assumed to be honest, i.e. it has no intention of starting a Man-In-the-Middle Attack (MIMA).

The steps towards secured social networking in the described scheme are as follows. If two users A and B want to securely share data through the proposed application, when the third party social networking application is started, its Application Programming Interfaces (APIs) are hooked with the described method, allowing users to encrypt data transparently. Whenever the application quits, original APIs are re-established so that user A can start communicating, but before sending any data, a secure channel needs to be set up by generating a session key. In order to do so a protocol integrating IBE and Diffie-Hellman key exchange is used. Both users A and B can then check whether the process was right by encrypting the peers ID and received messages using the session key. If one of the users is offline, a session key cannot be established so the initiator sets the session key by their choice.

As aforementioned, this scheme is not secure against a MIMA attack initiated by a malicious PKG, but this is supposed in the work that it will not to happen. Thanks to the use of IBE, non-PKG adversaries cannot start a MIMA attack. Furthermore, thanks to the use of Diffie-Hellman protocol, they neither can retrieve the session key.

The scheme proposed in that work includes a concrete application in which the described scheme is used. It was implemented for Android smartphones, using WeChat, and based on the Pairing-Based Cryptography library [18].

## 4    Proposal of NFC Authentication for IBE Communication

Bluetooth is a great technology, but experience so far has shown some of its weaknesses, which should be solved for a better use. One of its main drawbacks is the pairing process [19], because pairing two unknown devices requires over ten seconds, what makes that the user loses time in this operation. Furthermore, such a procedure should be transparent, letting the users focus on what they want to do and not on painful side operations. In conclusion, Bluetooth pairing lacks user-friendliness. In this regard, NFC is a breath of fresh air because it can be used for Bluetooth secure pairing [5]. In the following we propose a scheme using NFC-based Bluetooth pairing for providing a communication channel between smartphones, which is more secure than others described in previous works based on E0 or E1 stream ciphers [20]. In particular, the scheme here described applies IBE using phone numbers as public identities to deploy secure communication between devices.

Let us suppose that two users A and B want to securely share information through the Bluetooth of their smartphones. Assuming the existence of a TTP playing the role of the PKG server, which can register them as users of the system and provide them with corresponding private keys matching their identities, our protocol is divided into two main phases, whose main characteristics are show in Figures 1 and 2.
1. Setup phase. As soon as users tap their mobile phones, the pairing process begins. Following the NFC forum recommendation, smartphones transparently share their Bluetooth addresses, device names, and any other relevant information needed for the pairing process in an NFC Data Exchange Format (NDEF) message according to the Extended Inquiry Response (EIR) format specified in the Bluetooth Core Specification. Another important piece of information, the phone numbers, is also shared within this phase because they will be used as public identities later on in the scheme. Once finished this initial data transmission, the Bluetooth channel is already set up so that any information can be exchanged between the devices.

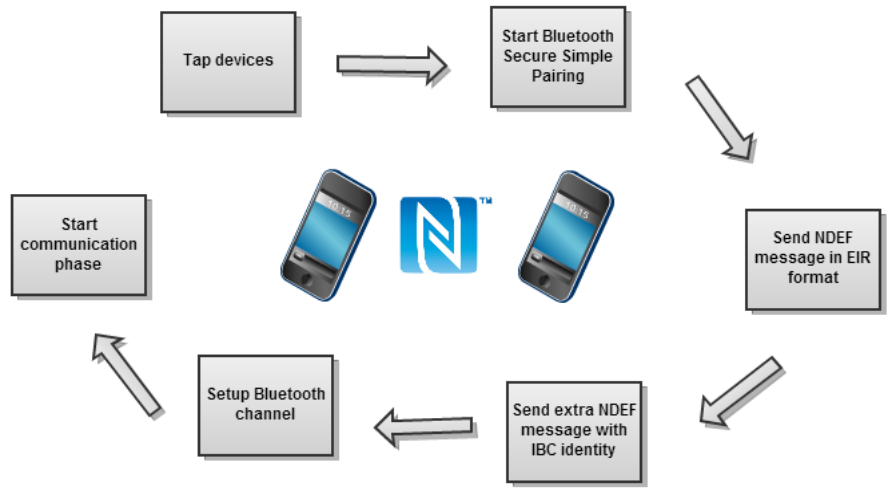

**Fig. 1.** Setup Phase

2. Communication phase. Once both devices have been paired, this stage begins. It is important to establish a real secure and trustworthy channel due to the known vulnerabilities of the Bluetooth cipher. IBC is a perfect approach to this problem so it is used in the proposal to guarantee confidentiality, integrity and authenticity. Depending on the size of the data to be shared, the initiating user can decide to send information directly by using an IBC or signcryption scheme, and/or other cryptographic schemes. For instance, in order to avoid a drastic increase in processing time, a session key is stated to be agreed for sharing large amounts of data through an efficient symmetric cipher such as AES by using an identity-based key agreement protocol like the ones used in [21, 22]. On the other hand, in order to prevent an increase in implementation complexity, an elegant but powerful Diffie-Hellman protocol is run by using the same IBC algorithm and by generating random keys through the inspection of data with smartphone sensors such as gyroscope, accelerometer, compass or GPS location, what allows obtaining enough randomness to establish a secure session key to be used in this communication phase.

The proposed protocol is simple, secure, transparent and extensible to be used with any other communication technologies such as Wi-Fi, so that the only necessary change is the NFC pairing detail during the setup phase.

In particular, the IBE algorithm used in the implementation of the communication phase basically follows the description of Boneh-Franklin (BF) scheme [2] based on bilinear pairings, with a few variations. Our implementation is based on a family of supersingular elliptic curves over finite fields $F_p$ of large prime characteristic p (between 512-bit and 7680-bit), known as type-1 curves. In the designed IBE, each user calculates its public key from its phone number, and the PKG calculates the corresponding private key from that public key by using its master key, as explained in the aforementioned extract step. In the encrypt step, the sender A chooses a

Content Encryption Key (CEK) and uses it to encrypt the message M, encrypts the CEK with the public key of the recipient B, and then sends to B both the encrypted message C and the encrypted CEK. In the decrypt step, B uses to decrypt the CEK its private key securely obtained from the PKG after a successful authentication, which is then used to decrypt the encrypted message C.

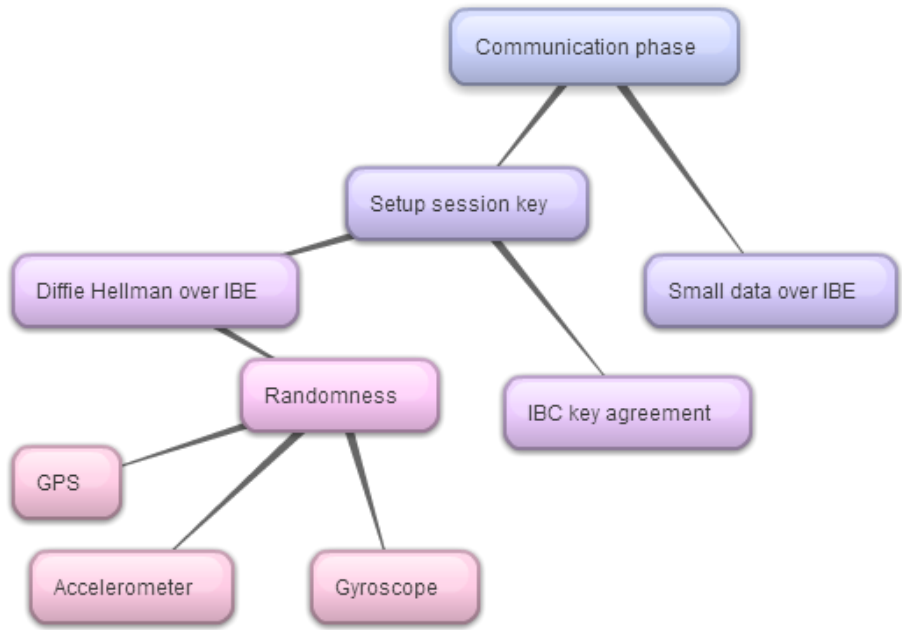

**Fig. 2.** Communication Phase

Regarding the elliptic curves used in the IBE schemes, the implementation of the application is based on type-1 curves E of the form $y^2=x^3+1$ defined over $F_p$ for primes p congruent to 11 modulo 12, mainly because these curves can be easily generated at random. The algorithm uses both the group $E(F_p)$ of points (x,y) satisfying the curve equation with affine coordinates x,y in $F_p$, and the group $E(F_p^2)$ of points (x,y) satisfying the curve equation with affine coordinates x,y in $F_p^2$, and with corresponding Jacobian projective coordinates X,Y,Z also in $F_p^2$.

Under the aforementioned conditions, the Tate pairing e takes as input two points P and Q in $E(F_p)$, and produces as output an integer in $F_p^2$. Thus, the Tate pairing is the core of the implemented IBE because it is a map that is hard to invert, but efficiently computable using Miller's algorithm and Solinas primes, thanks to its linearity in both arguments. However, when implementing it with smartphones, we found that although pairing in the projective system is faster than in the affine system, the cost is still very high in some cases, so pre-computation of some intermediate results, such as lambda for the sum of points, was used as solution to speed-up the whole process.

According to the implemented BF IBE, during the setup stage the PKG computes the master key as an integer s, and the public system parameters such as a prime p, a hash function h, the elliptic curve E, a point P in E($F_p$) and another point obtained by multiplying s times the point P. Afterwards, from the set of public parameters each public key is obtained by each user with phone number ID as a point $Q_{ID}$ in E($F_p$). During the extract step, the PKG returns to each applicant user with phone number ID, its private key in the form of another point $S_{ID}$ in E($F_p$). In this way, the encryption of a message M with the public key of a receiver whose phone number is ID, involves both a multiple rP of the point P (being r a randomly chosen integer) and the XOR of M and the hash of the r-power of the pairing result on both the public key $Q_{ID}$ and the public identity ID. In this way, the decryption of C is possible by adding to its second element, the hash of the pairing result of the private key and its first element.

The implementation of the proposed scheme has taken advantage of many primitives included in the pairing-based cryptographic library [8] because it is fully functional and ready-to-use. A few details concerning the technical features of the preliminary implementation of the proposed application are provided below.

The first implementation has been in the Windows Phone platform [23], and in particular in Windows Phone 8. We have used the Windows Phone 8 SDK, together with Visual Studio Ultimate 2012. All tests have been run either in the WP emulator or in a Nokia Lumia 920 smartphone. The specifications for the used smartphone are: processor is Qualcomm Snapdragon™ S4, processor type is Dual-core 1.5 GHz, 1 GB of RAM, and Bluetooth 3.0. The external libraries used in the Windows implementation have been: 32feet.NET.Phone, which is a shared source library to simplify Bluetooth development; and Bouncy Castle, which is a lightweight cryptography API for Java and C#. The average times obtained with the Nokia device are shown below:

| Message size (bytes) | Time to encrypt | Time to decrypt |
|---|---|---|
| 128 | 7497,198 ms | 7368,289 ms |
| 512 | 7498,221 ms | 6998,858 ms |

The implementation of the prototype is still being improved but, as shown above, the preliminary results obtained till now are promising.

## 5    Conclusions and Future Work

The general objective of this work has been the development of new ways to secure direct communications between smartphones, keeping them simple, efficient, energy-saving, functional and practical. Identity-Based Cryptography is perfect for that goal. Thus, the particular main aim of this paper has been to review known IBC-based protocols for mobile devices, in order to propose a new Bluetooth/Wi-Fi scheme using NFC for pairing and IBE for communications. Current work in progress is aimed at implementing the proposal in different testing environments and platforms, and at doing an in-depth analysis of its security. Our future research will focus on developing extensions such as signcryption and practical applications.

## Acknowledgements

Research supported by Spanish MINECO and European FEDER Funds under projects TIN2011-25452 and IPT-2012-0585-370000.

## References

1. Anson, A.: Smartphone Usage Statistics, ansonalex.com/infographics/smartphone-usage-statistics-2012-infographic, 2012.
2. Boneh, D., Franklin, M.K.: Identity-Based Encryption from the Weil Pairing, CRYPTO, LNCS 2139, pp. 213-229, 2001.
3. Hess, F.: Efficient identity based signature schemes based on pairings, ACM SAC, 2002.
4. Barreto, P.S.L.M., Libert, B., McCullagh, N., Quisquater J.-J.: Efficient and provably-secure identity-based signatures and signcryption from bilinear maps, ASIACRYPT, LNCS 3788, pp. 515-532, 2005.
5. NFC forum: www.nfc-forum.org.
6. Shamir, A.: Identity-Based Cryptosystems and Signature Schemes, CRYPTO, LNCS 7, pp. 47-53, 1984.
7. Boneh, D., Boyen, X., Shacham, H.: Short Group Signatures, CRYPTO, LNCS 3152, pp. 41-55, 2004.
8. Gentry, C., Silverberg, A.: Hierarchical ID-based cryptography, ASIACRYPT, LNCS 2501, pp. 548-566, 2002.
9. Boneh, D., Goh, E.-J., Boyen, X.: Hierarchical identity based encryption with constant size ciphertext, EUROCRYPT, LNCS 3493, pp. 440-456, 2005.
10. Bethencourt, J., Sahai, A., Waters, B.: Ciphertext-policy attribute based encryption, IEEE Symposium on S&P, 2007.
11. Goyal, V., Pandey, O., Sahai, A., Waters, B.: Attribute-based encryption for fine-grained access control of encrypted data, ACM CCS, 2006.
12. Chang, K., Zhang, X., Wang, G., Shin, K. G.: TGIS: Booting Trust for Secure Information Sharing in Dynamic Group Collaborations, IEEE PASSAT, pp. 1020-1025, 2011.
13. Jahid, S., Nilizadeh, S., Mittal, P., Borisov, N., Kapadia, A. : DECENT: A decentralized architecture for enforcing privacy in online social networks, IEEE PerCom, pp. 326-332, 2012.
14. Feldman, A.J., Blankstein, A., Freedman, M. J., Felten, E.W.: Social Networking with Frientegrity: Privacy and Integrity with an Untrusted Provider, USENIX Security, pp. 647-662, 2012.
15. Rahman, M.S., Huang, T.-K., Madhyastha, H.V., Faloutsos, M.: Efficient and Scalable Software Detection in Online Social Networks, USENIX Security, pp. 663-678, 2012.
16. Wu, Y., Zhao, Z., Wen, X.: Transparently Secure Smartphone-based Social Networking, IEEE WCNC, 2013.
17. Tencent: WeChat, www.wechat.com.
18. The Pairing Based Cryptography library: crypto.stanford.edu/pbc.
19. Barnickel, J., Wang, J., Meyer, U.: Implementing an Attack on Bluetooth 2.1+ Secure Simple Pairing in Passkey Entry Mode. IEEE TrustCom, 2012, pp. 17–24, 2012.
20. Yi, L., Vaudenay, S.: Cryptanalysis of Bluetooth Keystream Generator Two-Level E0. ASIACRYPT, pp. 483–499, 2004.

21. Zhang, L., Wu, Q., Qin, B., Domingo-Ferrer, J.: Provably secure one-round identity-based authenticated asymmetric group key agreement protocol, Information Sciences, 181(19): pp. 4318-4329, 2011.
22. Guo, H., Li, Z., Mu, Y., Zhang, X.: Provably secure identity based authenticated key agreement protocols with malicious private key generators, Information Sciences, 181(3): pp. 628-647, 2011.
23. NFC Secure Notes: www.windowsphone.com/s?appid=c72e51ce-fdda-452b-84c8-523cc27c76d5